\documentclass[letterpaper,showpacs,prb,preprint]{elsarticle}
\usepackage{epsfig}
\usepackage{float}
\usepackage{subcaption}
\usepackage{adjustbox}
\usepackage{array}
\usepackage{bm}
\usepackage{indentfirst}
\usepackage[textwidth=15cm]{geometry}
\usepackage{url}
\graphicspath{ {./} }
\newcolumntype{P}[1]{>{\centering\arraybackslash}p{#1}}

\begin{document}
	
	\begin{frontmatter}
		\title{First-principles investigation of low-dimension MSe$_2$ (M = Ti, Hf, Zr) configurations as promising thermoelectric materials}
		\author{Jonathan Tseng and Xuan Luo}
		\address{National Graphene Research and Development Center, Springfield, Virginia 22151, USA}
		\begin{abstract}
			\setlength{\parindent}{1cm}
			Interest in the application of thermoelectric devices for renewable energy has risen over the past decade. In this paper, we calculate the transport properties of various configurations of the transition metal dichalcogenide (TMD) MSe$_2$ (M = Hf, Zr, Ti) in search of promising thermoelectric materials at low/high temperatures. We explore the properties of the pure monolayer at discrete levels of biaxial tensile strain ($\epsilon = 0\%, 2\%, 4\%, 8\%$), as well as those of the van der Waals heterobilayers MSe$_2$/MSe$_2$ using first-principles calculations combined with semi-classical Boltzmann transport theory. It is found that all studied monolayers exhibit high thermoelectric performance at high temperatures, while the application of strain enhances Seebeck and thermopower non-linearly at low temperatures. The results also reveal the bilayer ZrSe$_2$/TiSe$_2$ to have remarkable thermopower at low temperatures. These findings offer insight into creating applications with enhanced performance at both low and high temperatures in future thermoelectric devices.
		\end{abstract}
	\end{frontmatter}
	\section{Introduction}
	Interest in renewable energy has steadily risen over the past decade, as demonstrated by the success of companies marketing alternative energy solutions\cite{boyle2004renewable}. Since almost two thirds of waste energy worldwide is lost as heat\cite{tan2016rationally}, significant interest has been observed in \textit{thermoelectric} and \textit{photo-thermoelectric} materials, which utilize heat generated by light or other sources to generate an electric current through the Seebeck effect\cite{chen2003recent}.
	
	Thermoelectric materials have been found to be of significant importance in industry, seeing use in energy harvesting\cite{tritt2006thermoelectric, zebarjadi2012perspectives, dehkordi2015thermoelectric}, solar/thermoelectric panel generators\cite{kraemer2011high}, and solid-state cooling systems\cite{rowe1995crc, goldsmid1986electronic, ziabari2016nanoscale}. Recent advances in the technology suggests also that sufficiently strong thermoelectric materials may be of use in photodetectors\cite{xu2009photo, freitag2013increased} through the photo-thermoelectric effect, potentially replacing conventional photovoltaic solutions. Thermoelectric performance is typically measured in terms of the dimensionless figure of merit $ZT = (\sigma S^2T/\kappa)$\cite{goldsmid2013thermoelectric}, where $\sigma$, S, T, and $\kappa$ represent the electrical conductivity, Seebeck coefficient, absolute temperature, and thermal conductivity, respectively. In order to operate at comparable efficiencies to conventional energy generation systems, thermoelectric materials of high ZT at both low and high temperatures are highly desirable\cite{vineis2010nanostructured}. To maximize ZT, researchers may seek to maximize power factor $(PF = S^2 \sigma)$ or minimize thermal conductivity $\kappa$. Improving the power factor has proven difficult in the past due to the physical coupling of the transport properties, which results in these goals conflicting. For instance, an increase in the electrical conductivity often results in a decrease in Seebeck coefficient S, and an increase in the electrical conductivity also leads to an increase in the thermal conductivity $\kappa$\cite{sadeghi2019non}. Subsequently, more efforts have been focused on the lowering of the thermal conductivity than the optimization of the power factor\cite{toberer2011phonon, biswas2012high}.
	
	Low-dimension materials have been given great interest due to the relative ease with which their electrical and transport properties can be tuned by simply modifying the number of layers. For example, reducing the thickness of the 1H-MX$_2$ (M = Mo, W; X = S, Se, Te) has been shown to promote greater tunability in the electronic band gap through a blue shift in the band gap energy\cite{kumar2012electronic}. Experiments have also confirmed that the PF of nanostructures can be much larger than in the bulk\cite{hicks1993effect, heremans2013thermoelectrics, ohta2007giant, hicks1996experimental}. Hicks and Dresselhaus proposed in 1993\cite{hicks1993effect} that 2D materials demonstrate high figures of merit due to quantum confinement effect, which causes sharp changes in the density of states (DOS) at band edges and subsequently much higher Seebeck coefficients than are found in the bulk. Increased DOS is also attainable through greater degeneracy in the electronic band structure\cite{mahan1998solid}; therefore, many attempts using various methods have been made to engineer the band structure of 2D materials\cite{pei2011convergence}. It has been generally shown that strain engineering is successful in the tuning of electronic and thermal transport properties\cite{ni2008uniaxial}. The application of strain shifts the bands, and in some cases, may cause a metal-semiconductor transition of states due to the opening of a new bandgap.
	
	Recently, the creation of stacked van der Waals heterostructures has emerged as a new class of materials. Among these, the transition metal dichalcogenide (TMD)-based heterostructures have been marked as very promising candidates for creating future thermoelectric materials -- computational studies have indicated that creating TMD-TMD heterostructures is effective for tuning the electronic and transport properties, and therefore the adjustment of the thermoelectric properties\cite{C3CP54080D, PhysRevB.87.075451, kou2013nanoscale}.
	
	1T-MSe$_2$ (M = Ti, Zr, Hf), henceforth referred to as MSe$_2$ for simplicity, is a semiconducting two-dimensional transition metal dichalcogenide (TMD) that has been shown to potentially exhibit low thermal conductivity\cite{ding2016thermoelectric} and a promisingly high figure of merit relative to its bulk counterpart\cite{ding2016thermoelectric, yan2019bilayer}. In this work, we perform a comprehensive examination of the transport properties of the MSe$_2$ monolayer, strained monolayer, and heterostructures in order to identify configurations that optimize thermoelectric transport qualities, particularly the power factor (PF), at both low and high temperatures. Therefore, we investigate these materials in terms of S, $\sigma$, and PF.

	\section{Method}	
	
	We studied strained and unstrained monolayer MSe$_2$ (M = Ti,Zr,Hf), as well as the stacked heterobilayers  of the form MSe$_2$ / MSe$_2$. MSe$_2$ layers crystallize in the 1T-CdI$_2$ structure, consisting of two planes of Se separated by a hexagonal close-packed plane of M (M = Ti, Zr, Hf), as shown in Figure \ref{structure}.
	
	\subsection{Computational Details}
	
	Density functional theory (DFT)\cite{hohenberg1964inhomogeneous, kohn1965self} calculations were
	performed with ABINIT\cite{gonze2005, gonze2009} using generalized gradient
	approximation (GGA) with Perdew-Burke-Ernzerhof (PBE) functionals\cite{perdew1996}. We choose projector augmented wave (PAW)\cite{paw, torrent2008implementation, blochl1994projector} pseudopotentials generated
	with AtomPAW\cite{holzwarth2001projector}. The electron configuration and radius cutoff for the pseudopotentials of each element used in our
	calculations are shown in Table \ref{radcut}.
	
	\begin{table}[hpt]
		\caption{The valence electron configuration and radius cutoff used for generating PAW pseudopotentials}
		\centering
		\begin{tabular}{c c c}
			Element&Valence Configuration&Radius Cutoff (a.u)\\
			\hline
			\hline
			Ti&$3s^2 3p^6 4s^1 3d^3$&2.3\\
			Zr&$4s^2 4p^6 5s^1 4d^3$&2.21\\
			Hf&$5s^2 5p^6 6s^2 5d^2$&2.41\\
			Se&$4s^2 4p^4$&2.2\\
		\end{tabular}
		\label{radcut}
	\end{table}
	
	\subsection{Convergence and Relaxation}
	To balance computational time and accuracy, we converged the kinetic energy cutoff and k-point mesh parameters of all materials. The tolerance for self-consistent field (SCF) cycles was set to $1.0\times10^{-10}$ Ha, and each parameter was declared converged when the energy difference between consecutive datasets was less than $1.0\times10^{-4}$ Ha (about 3 meV) twice successively. The energy cutoff and k-mesh we used for each material can be found in Table \ref{ktable}.
	
	Using the converged kinetic energy cutoff and k-point mesh, we performed Broyden-Fletcher-Goldfarb-Shanno (BFGS)\cite{broyden1970convergence, fletcher1970new, goldfarb1970family, shanno1970conditioning, shanno1970optimal} structural optimization to determine the lattice parameters and atomic positions of each compound. Each SCF cycle terminated when the difference between consecutive forces was less than $1.0 \times 10^{-6}$ Ha/Bohr. The relaxation finished when the maximal absolute force on each atom was less than $5.0 \times 10^{-5}$ Ha/Bohr. 
	
	We used the large value of 30 Bohr\cite{khan2019theoretical,sadeghi2019non, guo2016biaxial} as the vacuum size for monolayer and bilayer calculations, and employed an inter-layer distance greater than 6 Bohr\cite{khan2019theoretical} in the case of the bilayer to minimize interactions between adjacent layers. 
	
	\begin{table}[h!]
		\centering
		\caption{The converged kinetic energy cutoffs and k-point meshes for each material}
		\begin{tabular}{c  c c}
			Material&Energy cutoff (Ha)&k-mesh\\
			\hline
			\hline
			TiSe$_2$&15&12$\times$12$\times$1\\
			HfSe$_2$&23&10$\times$10$\times$1\\
			ZrSe$_2$&17&10$\times$10$\times$1\\
			TiSe$_2$/HfSe$_2$&17&12$\times$12$\times$1\\
			TiSe$_2$/ZrSe$_2$&17&12$\times$12$\times$1\\
			HfSe$_2$/ZrSe$_2$&17&12$\times$12$\times$1\\
		\end{tabular}
		\label{ktable}
	\end{table}
	
	These converged parameters were used to calculate the electronic band structure and transport properties of each material.
	
	\subsection{Band Structure}
	
\begin{figure} [H]
		\centering
		\begin{subfigure}[b]{0.45\textwidth}
			\includegraphics[width=\textwidth]{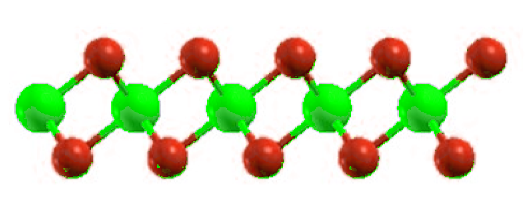}
			\caption{}
			\label{structure1}
		\end{subfigure}
		\begin{subfigure}[b]{0.45\textwidth}
			\includegraphics[width=\textwidth]{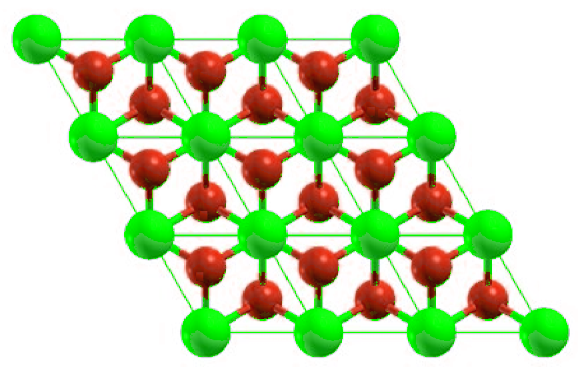}
			\caption{}
			\label{structure2}
		\end{subfigure}
		\begin{subfigure}[b]{0.45\textwidth}
			\includegraphics[width=\textwidth]{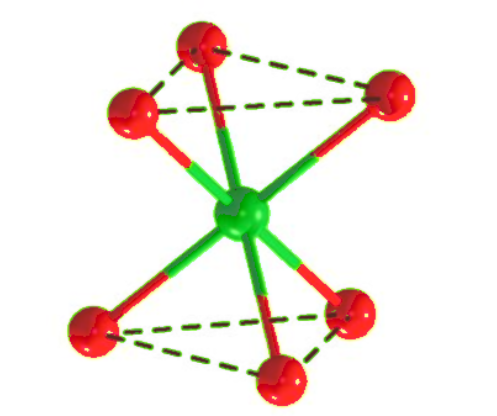}
			\caption{}
			\label{structure3}
		\end{subfigure}
		\begin{subfigure}[b]{0.45\textwidth}
			\includegraphics[width=\textwidth]{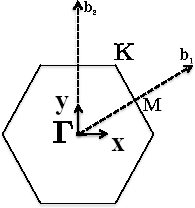}
			\caption{}
			\label{kpath}
		\end{subfigure}
		\caption{The 1T-$MSe_2$ $M = (Ti,Hf,Zr)$ structure shown (a) from the side, (b) from above (unit cell outlined), and (c) in perspective. Green = M, Red = Se. (d) The first Brillouin Zone, with high-symmetry k-points $\Gamma$, M, and K shown.}
		\label{structure}
	\end{figure}

	We used the k-points $\Gamma $ (0.0, 0.0, 0.0), M ($1/2$, 0.0, 0.0), and K ($1/3$, $1/3$, 0.0), the points that are in the plane of the layer (Figure \ref{kpath}) for monolayer and bilayer calculations. The k-point sampling was developed using the shift (0.0, 0.0, 0.5).
	
	\subsection{Transport Properties}
	Transport properties were generated using the solution of the semi-classical Boltzmann transport equation within the constant relaxation time approximation (CRTA) used by the program BoltzTraP2 \cite{BoltzTraP2}. The CRTA method has been used many times to successfully predict the transport properties of materials and has been shown to have adequate predictive power and accuracy\cite{sadeghi2019non, khan2019theoretical}. Under the CRTA, the electrical conductivity and Seebeck coefficient are expressed as
	
	\begin{equation}
	{[\sigma]_{i j}(\mu, T)=e^{2} \int_{-\infty}^{+\infty} d E\left(-\frac{\partial f(E, \mu, T)}{\partial E}\right) \Sigma_{i j}(E)} \\
	\label{Equation1}
	\end{equation}
	\begin{equation}
	{[\sigma S]_{i j}(\mu, T)=\frac{e}{T} \int_{-\infty}^{+\infty} d E\left(-\frac{\partial f(E, \mu, T)}{\partial E}\right)(E-\mu) \Sigma_{i j}(E)} \\
	\label{Equation2}
	\end{equation}	
	
	where f is the Fermi-Dirac distribution, e$(<0)$ is the electron charge, and T is the temperature. $\Sigma_{i j}(E)$ is the transport distribution function
	
	\begin{equation}
	\Sigma_{i j}(E)=\frac{1}{V N_{k}} \sum_{n, \mathbf{k}} v_{i}(n, \mathbf{k}) v_{j}(n, \mathbf{k}) \tau_{n, \mathbf{k}} \delta\left(E-E_{n, \mathbf{k}}\right)
	\label{Equation5}
	\end{equation}
	
	where V is the volume of the unit cell, $E_{n,\mathbf{k}}$ is the energy of an electron in the n-th band in wave vector k, $v_{i}(n, \mathbf{k})$ is the i-th cartesian component of its velocity, and the summation is conducted over the whole Brillouin Zone.

	\section{Results and Discussion}
	In the following section we present our converged lattice parameters as compared to previous theoretical and experimental results, as well as an analysis of the calculated band structures and transport properties of low-dimension MSe$_2$ configurations.
	
	\subsection{Lattice Parameters}
	
	We performed BFGS optimization on each monolayer's unit cell. The converged lattice parameters as compared to previous experimental and theoretical studies can be found in Table \ref{vac}.
	
\begin{figure} [H]
		\centering
		\begin{subfigure}[b]{0.3\textwidth}
			\centering
			\includegraphics[height=5cm]{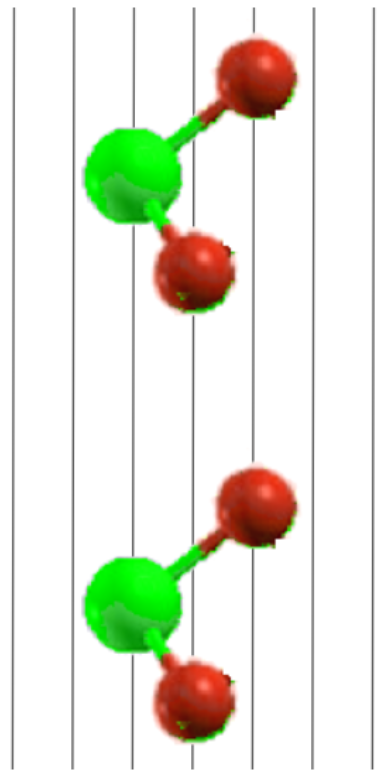}
			\caption{}
			\label{config1}
		\end{subfigure}
		\begin{subfigure}[b]{0.3\textwidth}
			\centering
			\includegraphics[height=5cm]{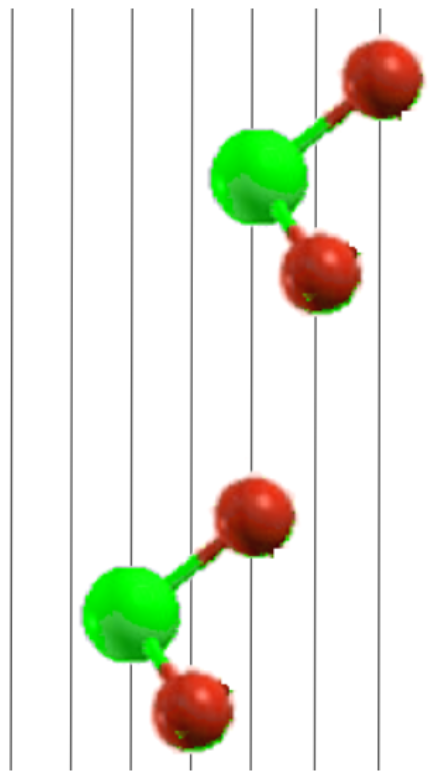}
			\caption{}
			\label{config2}
		\end{subfigure}
		\begin{subfigure}[b]{0.3\textwidth}
			\centering
			\includegraphics[height=5cm]{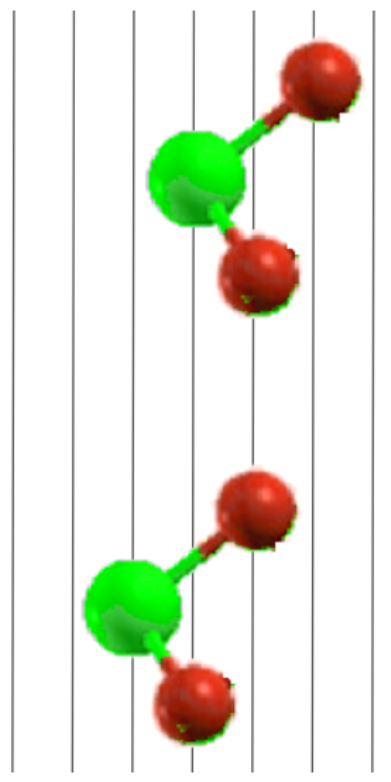}
			\caption{}
			\label{config3}
		\end{subfigure}
		\caption{The considered bilayer stacking configurations, where (a) Transition metal is located on top of transition metal, (b) Transition metal is located on top of chalcogen atom (top layer), (c) Transition metal is located on top of chalcogen atom (bottom layer) }
		\label{configs}
	\end{figure}

	In heterobilayers, monolayers are stacked on top of each other in 2-layer supercells. Because the lattice mismatch between the constituent monolayers was small, the lattice parameters were simply re-averaged before stacking.  Since the electronic structure is sensitive to layer stacking configuration, different options were explored: a) transition metal on top of transition metal, b) transition metal on top of chalcogen (top layer), and c) transition metal on top of chalcogen (bottom layer); see Figure \ref{configs}. The total energies of each configuration were compared, with the lowest indicating the most stable configuration. Results showed that configuration (c) was preferred; therefore, all bilayer calculations were performed with this configuration.
	
	\begin{table}[hpt]
		\caption{Lattice parameters, as calculated in current study vs. previous experimental and theoretical results}
		\resizebox{\textwidth}{!}{
		\begin{tabular*}{\textwidth}{P{2cm} P{3cm} P{3cm} P{3cm} P{2cm}}
			Material&Current Study (Bohr)&Experimental results (Bohr)&Theoretical results (Bohr)&Error (\%)\\
			\hline
			\hline
			TiSe$_2$&6.779&6.689$\pm$0.113\cite{zheng2018electrical,peng2015molecular}&6.520\cite{sadeghi2019non}&1.34\%\\
			HfSe$_2$&7.128&7.078\cite{hodul1984anomalies}&7.086\cite{khan2019theoretical}&0.71\%\\
			ZrSe$_2$&7.178&7.120\cite{le1989synthesis}&7.143\cite{khan2019theoretical}&0.82\%\\\\
		\end{tabular*}}
		\label{vac}
	\end{table}

	The converged lattice parameters agree closely with experimental and theoretical data, with the only significant difference found in the parameter of TiSe$_2$ as compared to previous theoretical results. This can be explained by a difference in the functional used: the study in question used the GGA-PBESOL functional.
	
	\subsection{MSe$_2$ monolayer}
	
\begin{figure} [H]
		\centering
		\begin{subfigure}[b]{0.3\textwidth}
			\includegraphics[width=\textwidth]{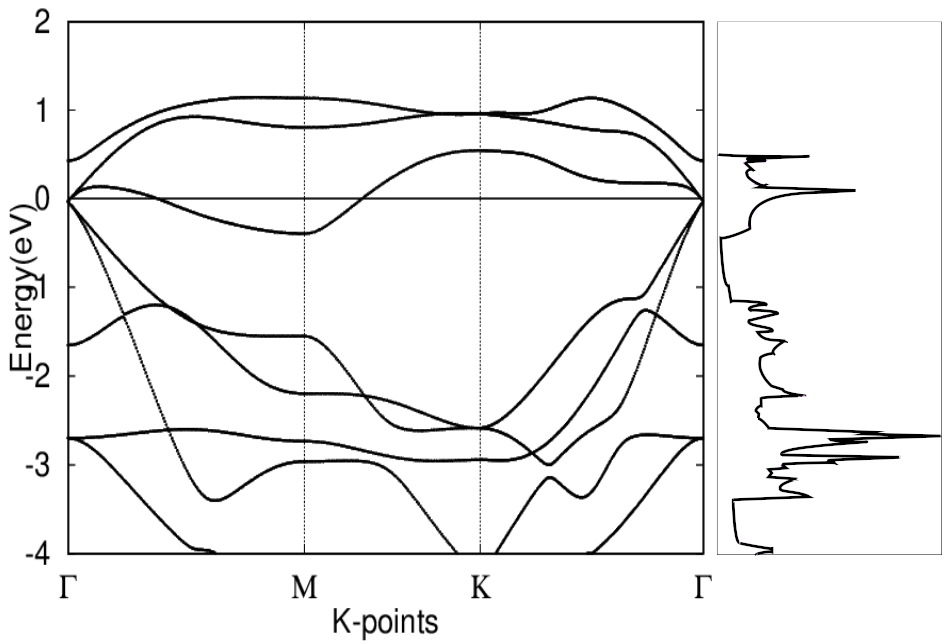}
			\caption{}
			\label{Tiband}
		\end{subfigure}
		\begin{subfigure}[b]{0.3\textwidth}
			\includegraphics[width=\textwidth]{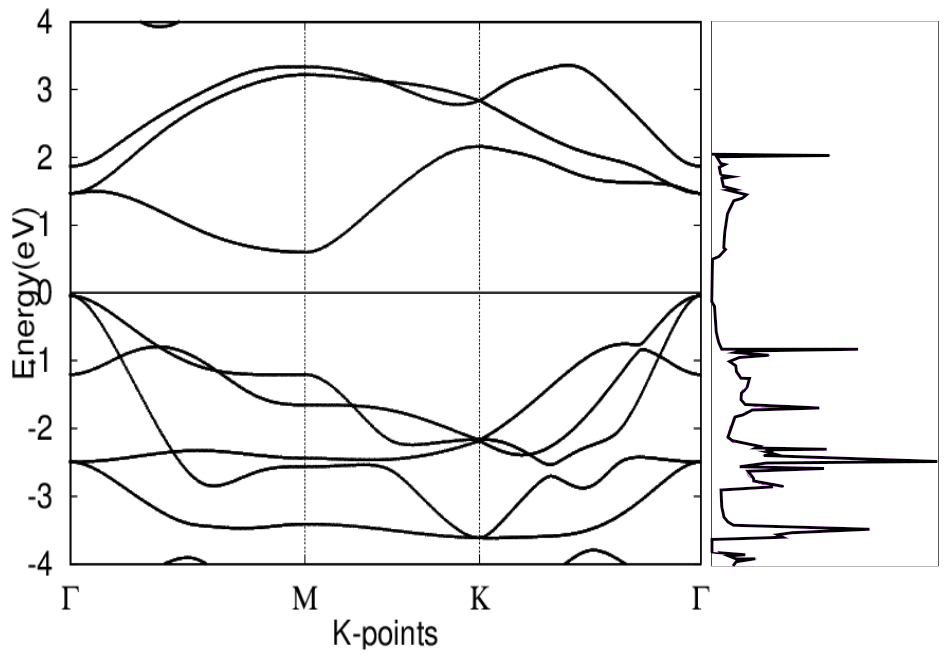}
			\caption{}
			\label{Hfband}
		\end{subfigure}
		\begin{subfigure}[b]{0.3\textwidth}
			\includegraphics[width=\textwidth]{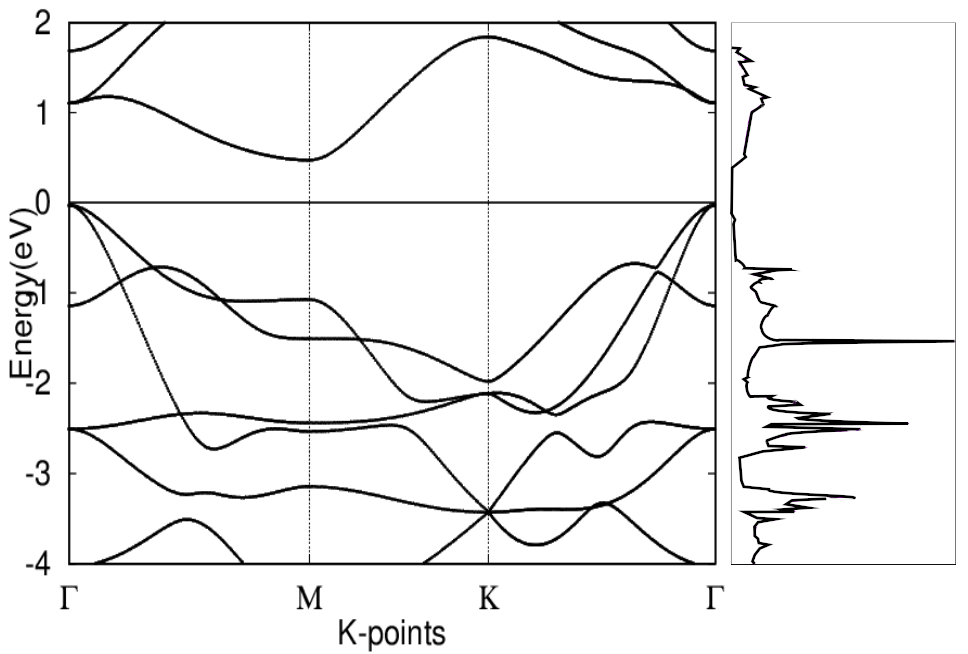}
			\caption{}
			\label{Zrband}
		\end{subfigure}
		\caption{The electronic band structure and DOS calculated at the GGA-PBE level for monolayer (a) TiSe$_2$, (b) HfSe$_2$, and (c) ZrSe$_2$}
	\end{figure}

	Figures \ref{Tiband}, \ref{Hfband}, and \ref{Zrband} show the electronic band structures of HfSe$_2,$ ZrSe$_2$, and TiSe$_2$ monolayers at the GGA-PBE level calculated along high-symmetry directions of the first Brillouin zone, shown in Figure \ref{kpath} 
	
	Band structures show that HfSe$_2$ and ZrSe$_2$ are indirect bandgap semiconductors with valence band maximum (VBM) at $\Gamma$- point and conduction band minimum (CBM) at M-point.
	
	Notably, a band overlap of about 0.327 eV can be observed in the structure of TiSe$_2$. This result agrees with previous theoretical DFT studies \cite{chen2015charge, singh2017stable}, but disagrees with experimental studies, which indicated through angle-resolved photoemission spectroscopy that TiSe$_2$ exhibited a bandgap of about 98 meV\cite{chen2015charge, sugawara2015unconventional}. This discrepancy is best explained by the tendency of DFT calculations to under-estimate bandgaps\cite{johari2012tuning}.	
	
	Previous research has reported that hybrid functional calculations agree better with experiments than semi-local functionals. However, this conclusion cannot be generalized, and has been demonstrated to vary by material \cite{johari2012tuning}. The consistent behavior of the PBE functional leads us to choose it over hybrid functionals.
	
\begin{figure} [H]
		\centering
		\includegraphics[width=0.7\textwidth]{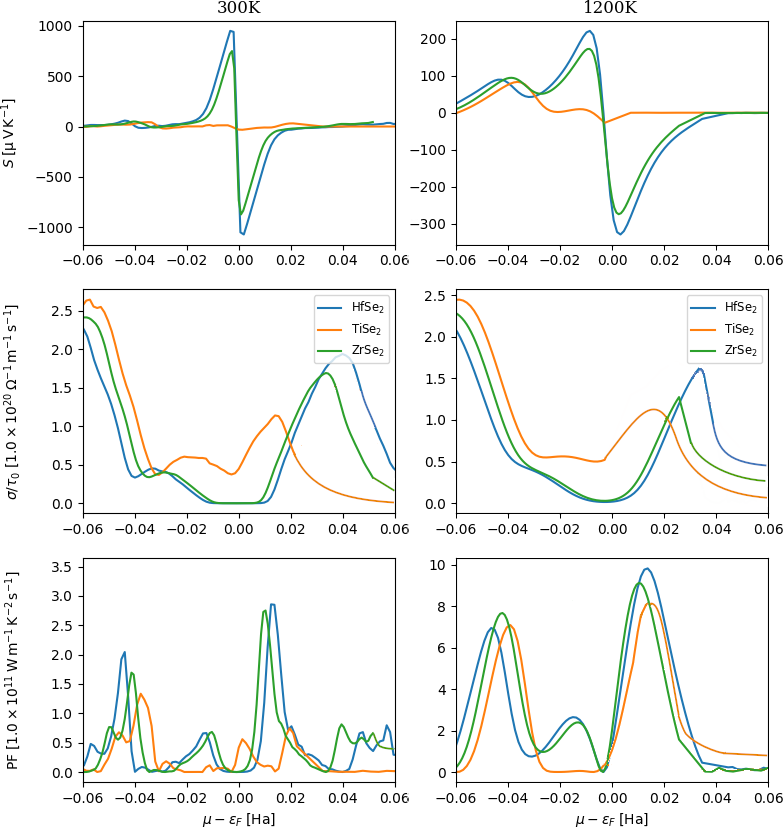}
		\label{Monotransport1}
		\caption{Thermoelectric properties: Seebeck coefficient (S), electrical conductivity ($\sigma/\tau$), power factor (PF) plotted as a function of chemical potential at 300K and 1200K for monolayer HfSe$_2$, ZrSe$_2$, and TiSe$_2$}
		\label{Monotransport}
	\end{figure}

	Transport coefficients as a function of chemical potential $\mu$ of the compound are calculated by solving Equations 1, 2, and 3. Chemical potential varies with the doping of a compound: $\mu$ is positive in the n-type doping region, and $\mu$ is negative in the p-type doping region. The results, calculated near room temperature (300K) and at high temperature (1200K), are shown in Figure \ref{Monotransport} and Table \ref{transporttable}. 
	
	At both 300K and 1200K, significantly higher values of Seebeck coefficient S are observed around $\mu = 0$, revealing that greatest S can be achieved with small doping. Generally, “promising” thermoelectric materials are presented as having S of more than $200$ $\mu$V/K \cite{lv2016strain,C7RA08828K}. At 300K, peak S values are $1070$, $873$, and $43.7$ ($\mu$V/K) for HfSe$_2$, ZrSe$_2$, and TiSe$_2$, respectively. HfSe$_2$ and ZrSe$_2$ are immediately seen to be of great promise, while TiSe$_2$ shows significantly lower Seebeck than the other monolayers due to its metallic (low bandgap) nature. The high S in HfSe$_2$ and ZrSe$_2$ is reflected by sharp peaks and valleys in the DOS at band edges, confirming studies that valley degeneracy is correlated with Seebeck \cite{pei2011convergence}. At 1200K, significantly lower S is observed for semiconductor monolayers, at about $1/4$ of the value at 300K, but TiSe$_2$ exhibits increased Seebeck at higher temperatures, which is consistent with previous theoretical studies\cite{sadeghi2019non}.
	
	Electrical conductivity divided by constant relaxation time ($\sigma/\tau$) for all materials does not differ much with temperature, showing much lower temperature sensitivity than Seebeck. At both 300K and 1200K, a higher conductivity is seen in p-type region than in n-type region, and very nearly the same trend in conductivity can be seen. Electrical conductivity is shown to be inversely related to Seebeck; whereas Seebeck shows greatest value at low chemical potential, $\sigma$ shows greatest increase at high chemical potential. Peak $\sigma$ values show a minor decrease, and shift slightly towards lower potential values. TiSe2 exhibits higher peak conductivity and less conductivity variance with $\mu$ due to its smaller bandgap nature.

	To gauge thermoelectric performance, power factor (PF) is also shown. PF is shown to be sensitive to temperature due to its Seebeck component. At both 300K and 1200K, a greater PF is seen in the n-type region than the p-type region because of greater Seebeck. Notably, while both Seebeck and electrical conductivity were shown to decrease with temperature, PF shows greater values at 1200K than 300K due to greater electrical conductivity in the region. The PF values of all monolayers (both at 300K and 1200K) are notably high compared to other previously investigated low-dimension materials\cite{Dimple_2017}, making for potential applications in thermoelectric devices. Further, HfSe$_2$ shows PF greater than the other materials, showing greatest thermoelectric promise among these pure monolayers, especially at high temperatures. 
	
	\begin{table}[hpt]
		\caption{Calculated peak transport values: Seebeck coefficient (S), electrical conductivity ($\sigma/\tau$), power factor (PF) for MSe$_2$ monolayers at 300K and 1200K}
		\centering
		\begin{tabular}{c c c c}
			Material&S ($\mu$V/K)& $\sigma/\tau$ (10$^{-4}$ $\Omega^{-1}m^{-1}s^{-1}$)&PF (10$^{11}$ W/K$^2$ms)\\
			\hline
			\hline
			HfSe$_2$ (300K)&1070&2.42&2.92\\
			ZrSe$_2$ (300K)&873&2.41&2.75\\
			TiSe$_2$ (300K)&43.7&2.64&1.33\\
			HfSe$_2$ (1200K)&329&2.25&9.81\\
			ZrSe$_2$ (1200K)&274&2.31&9.11\\
			TiSe$_2$ (1200K)&83.7&2.44&8.08\\
		\end{tabular}
		\label{transporttable}
	\end{table}
	
	\subsection{Strained MSe$_2$ monolayer}
	
\begin{figure} [H]
		\centering
		\begin{subfigure}[b]{0.3\textwidth}
			\includegraphics[width=\textwidth]{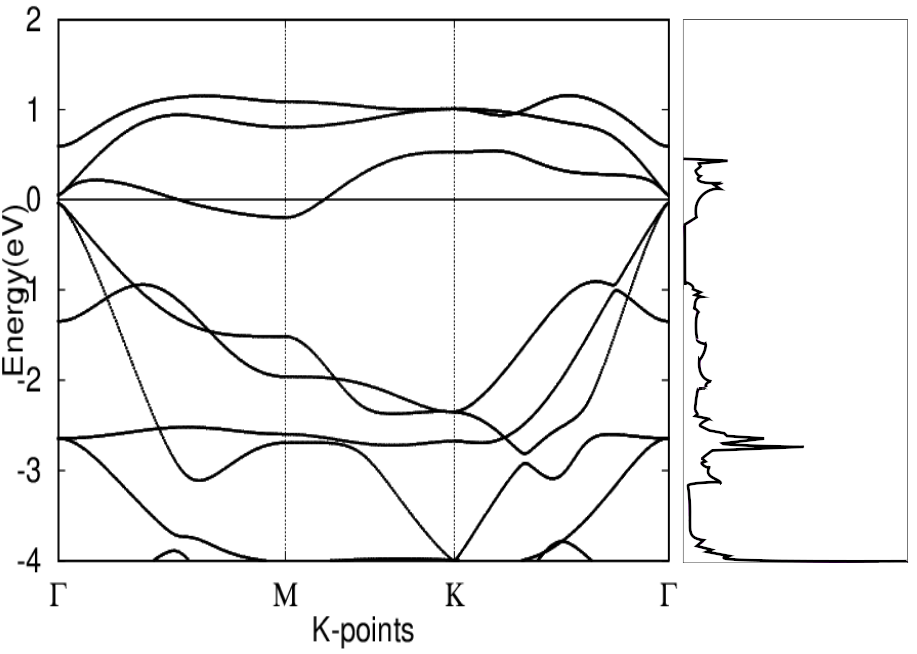}
			\caption{}
			\label{T:1}
		\end{subfigure}
		\begin{subfigure}[b]{0.3\textwidth}
			\includegraphics[width=\textwidth]{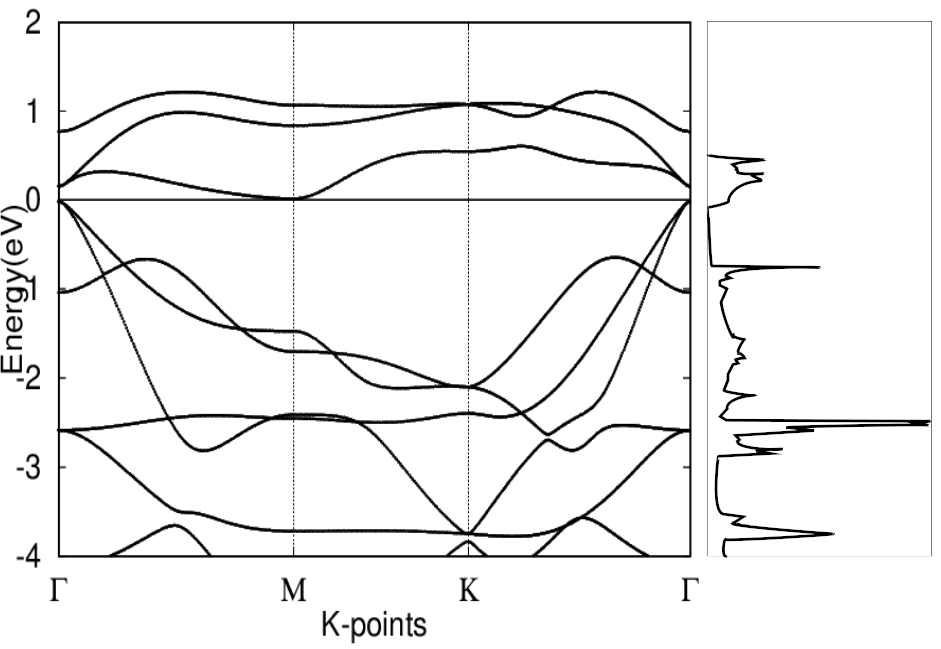}
			\caption{}
			\label{T:2}
		\end{subfigure}
		\begin{subfigure}[b]{0.3\textwidth}
			\includegraphics[width=\textwidth]{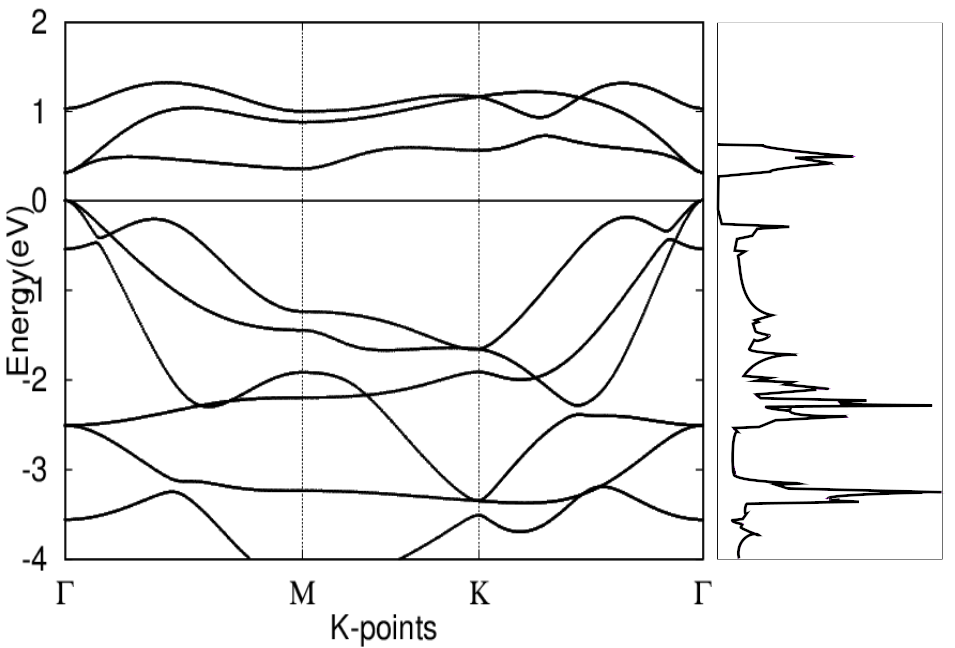}
			\caption{}
			\label{T:3}
		\end{subfigure}
		\caption{Band structure and DOS calculated for TiSe$_2$ under tensile strain: $\epsilon$ = (a) 2\% (b) 4\% (c) 8\%}
		\label{Tistrainband}
	\end{figure}

\begin{figure} [H]
		\centering
		\begin{subfigure}[b]{0.3\textwidth}
			\includegraphics[width=\textwidth]{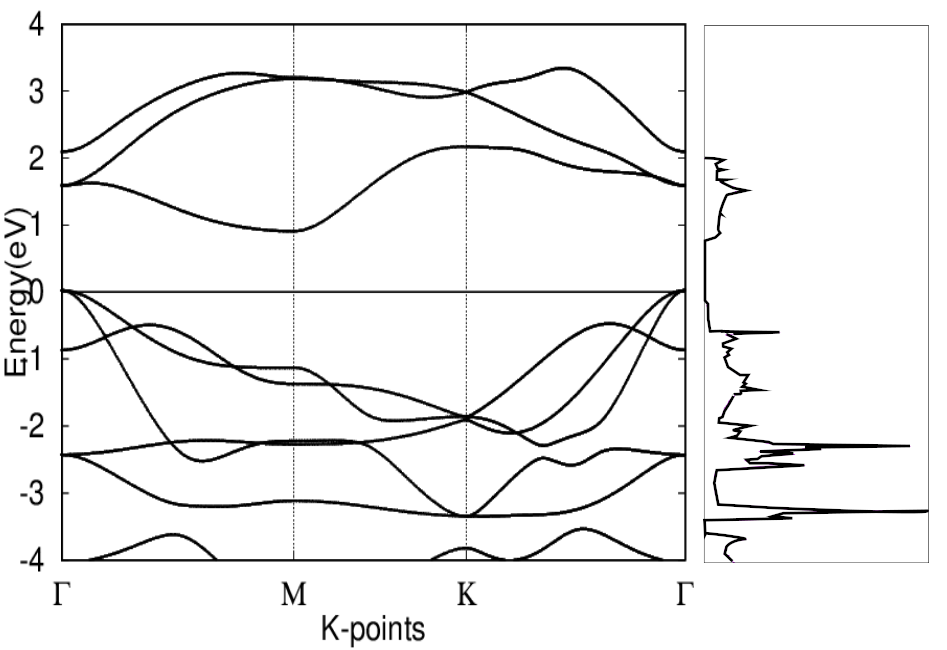}
			\caption{}
			\label{H:1}
		\end{subfigure}
		\begin{subfigure}[b]{0.3\textwidth}
			\includegraphics[width=\textwidth]{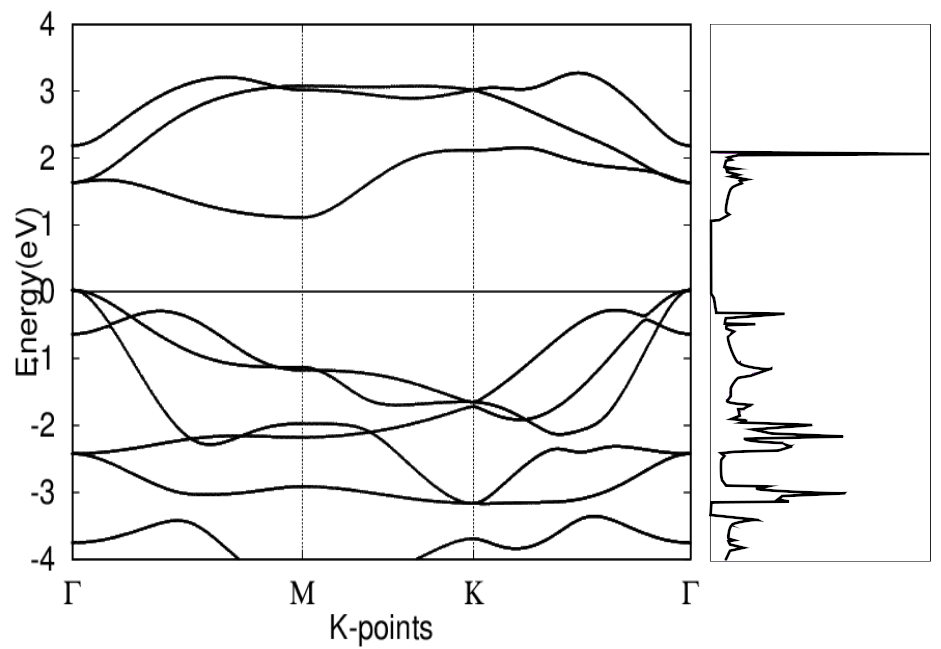}
			\caption{}
			\label{H:2}
		\end{subfigure}
		\begin{subfigure}[b]{0.3\textwidth}
			\includegraphics[width=\textwidth]{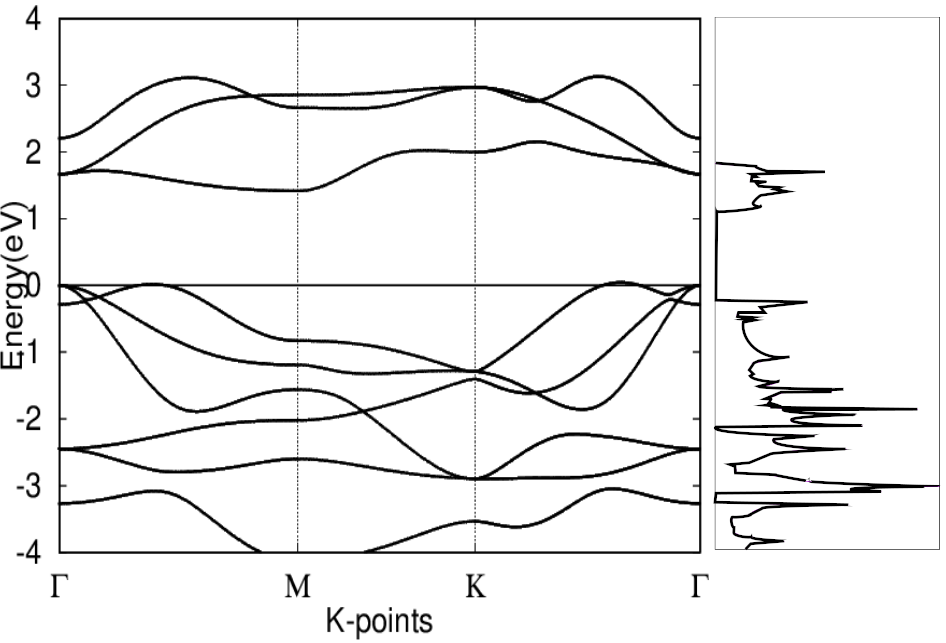}
			\caption{}
			\label{H:3}
		\end{subfigure}
		\caption{Band structure and DOS of HfSe$_2$ under tensile strain: $\epsilon$ = (a) 2\% (b) 4\% (c) 8\%}
		\label{Hfstrainband}
	\end{figure}

\begin{figure} [H]
		\centering
		\begin{subfigure}[b]{0.3\textwidth}
			\includegraphics[width=\textwidth]{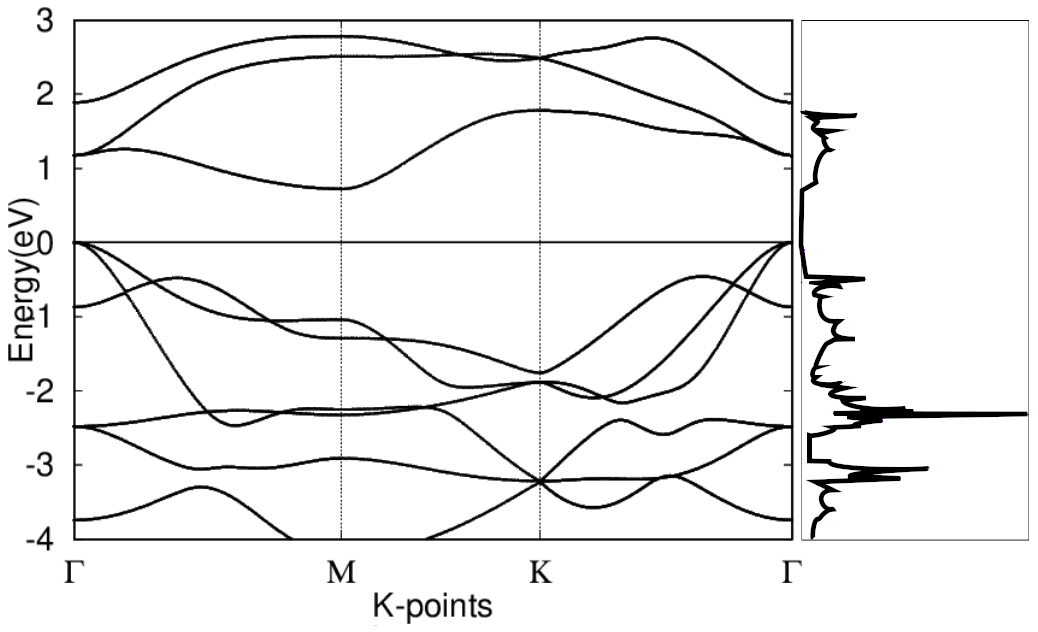}
			\caption{}
			\label{Z:1}
		\end{subfigure}
		\begin{subfigure}[b]{0.3\textwidth}
			\includegraphics[width=\textwidth]{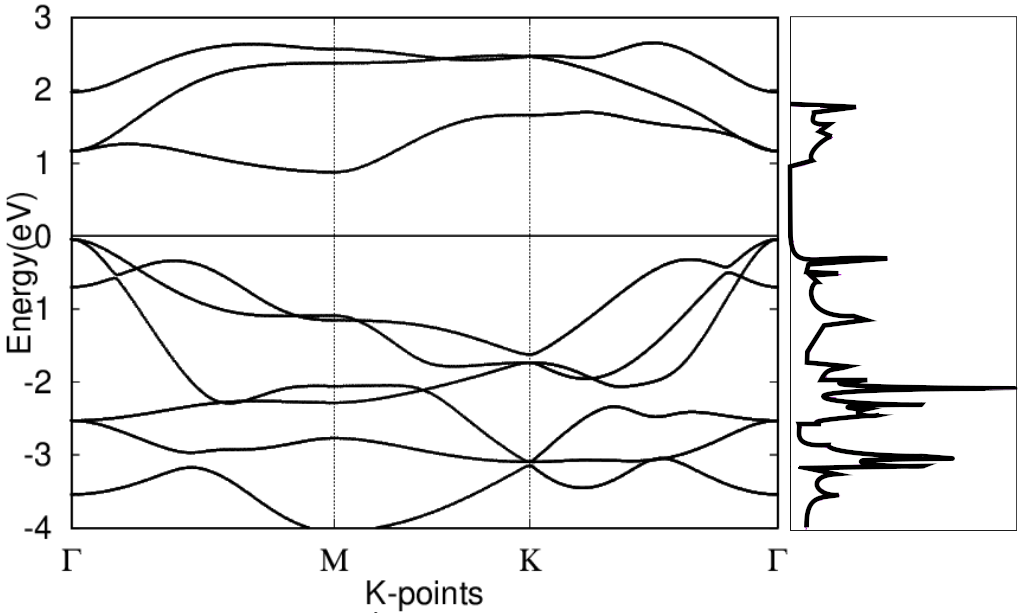}
			\caption{}
			\label{Z:2}
		\end{subfigure}
		\begin{subfigure}[b]{0.3\textwidth}
			\includegraphics[width=\textwidth]{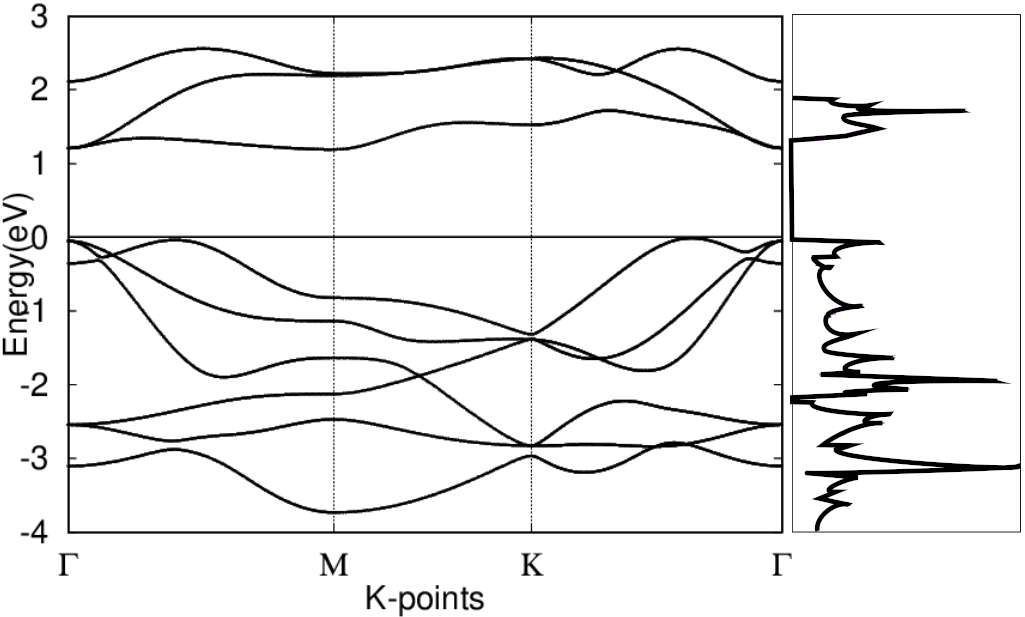}
			\caption{}
			\label{Z:3}
		\end{subfigure}
		\caption{Band structure and DOS of ZrSe$_2$ under tensile strain: $\epsilon$ = (a)2\% (b)4\% (c)8\%}
		\label{Zrstrainband}
	\end{figure}

	Strain is known to be an effective way to tune the electronic properties of 2D materials \cite{pei2011convergence, guo2016biaxial, guo2017small, lv2016strain}. Tensile strain leads to a decreased bandwidth and therefore an increase in the bandgap. To study the effect of tensile strain, uniform biaxial strains of $\epsilon$ = 2\%, 4\%, and 8\% were applied to each material. For each strain, atomic positions were relaxed in the out-of-plane direction. Under tensile strain, the Se atoms move closer to the M atom layer. The band structures of each material under strain are shown in Figures \ref{Tistrainband}, \ref{Hfstrainband}, and \ref{Zrstrainband}.
	
	The application of tensile strain is observed to consistently widen the bandgap in all monolayers, progressively flattening the conduction bands as strain increases. Tensile strain  is also observed to increase the number of peaks and valleys in the DOS of all materials, most noticeably near the fermi level at the edges of the bandgap.
	
	To examine the thermoelectric properties of each material, the power factor, electrical conductivity, and Seebeck coefficient were calculated as a function of carrier concentration using Equations \ref{Equation1} and \ref{Equation2}.
	
	\begin{figure} [H]
		\centering
		\includegraphics[width=\textwidth]{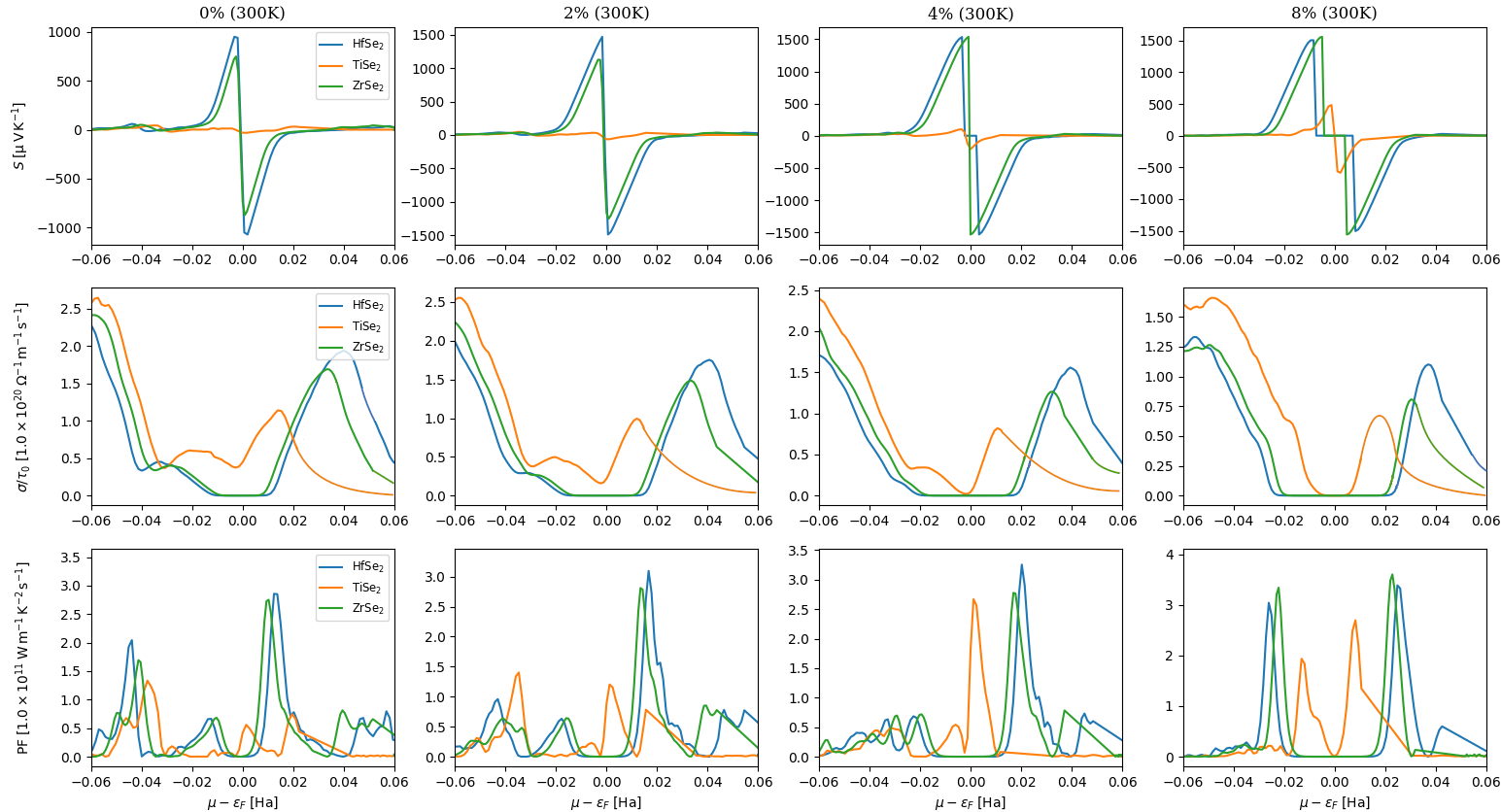}
		\includegraphics[width=\textwidth]{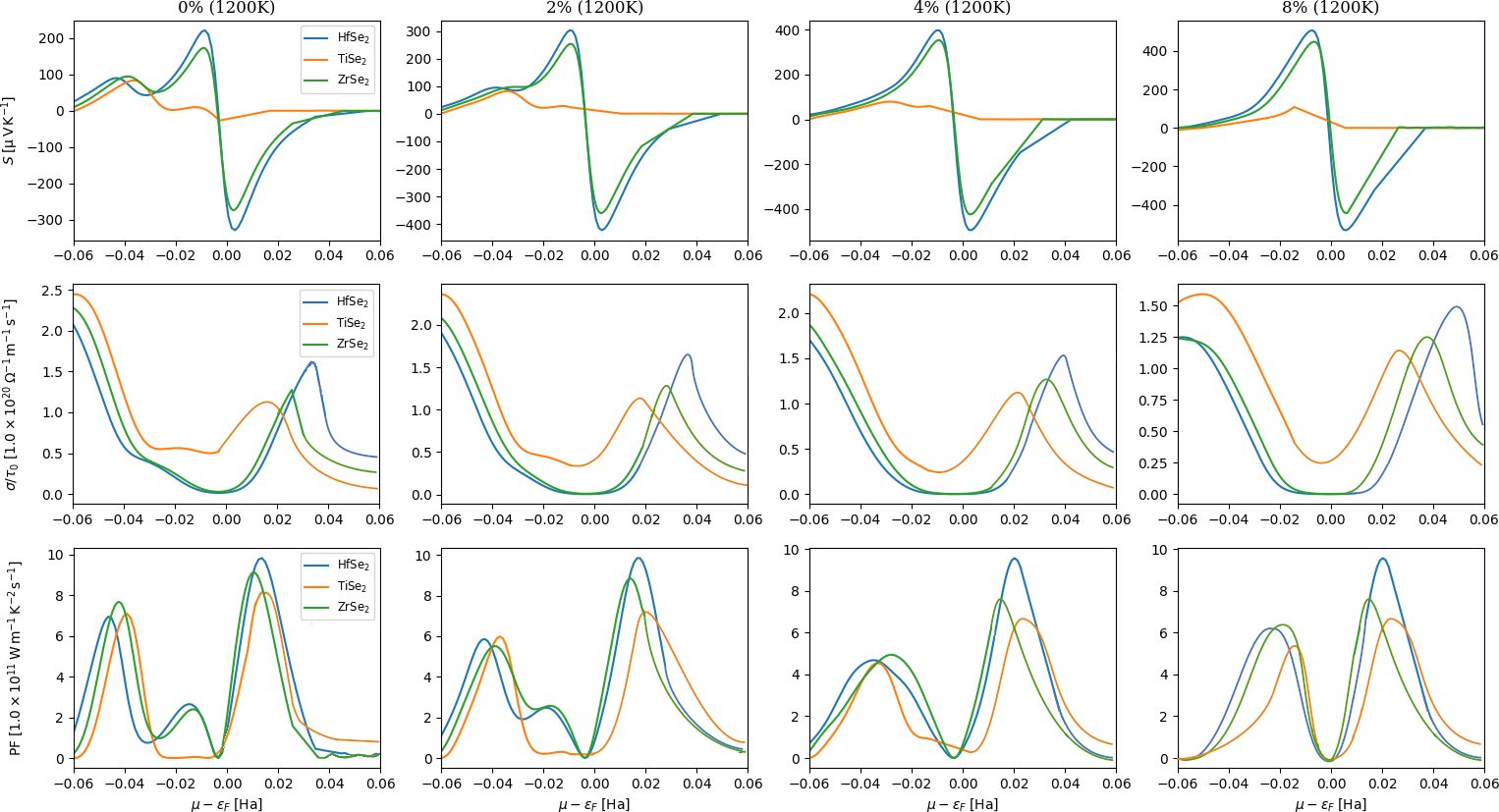}
		\label{Straintransport}
		\caption{Thermoelectric properties: Seebeck coefficient (S), electrical conductivity ($\sigma/\tau$), power factor (PF) plotted as a function of chemical potential at 300K (top 3 rows) and 1200K (bottom 3 rows) with varying tensile strains ($\epsilon$ = 2\%, 4\%, 8\%) for: monolayer HfSe$_2$, ZrSe$_2$, and TiSe$_2$} 
	\end{figure}
	\begin{table}[H]
		\caption{Calculated peak transport values: Seebeck coefficient (S), electrical conductivity ($\sigma/\tau$), power factor (PF) for strained MSe$_2$ monolayers at 300K and 1200K}
		\centering
		\begin{tabular}{c c c c}
			Material&S ($\mu$V/K)&$\sigma/\tau$ (10$^{-4}$ $\Omega^{-1}m^{-1}s^{-1}$)&PF (10$^{11}$ W/K$^2$ms)\\
			\hline
			\hline
			HfSe$_2$ (300K)&1070&2.42&2.92\\
			HfSe$_2$ 2\% (300K)&1490&2.06&3.13\\
			HfSe$_2$ 4\% (300K)&1530&1.70&3.34\\
			HfSe$_2$ 8\% (300K)&1500&1.33&3.40\\
			ZrSe$_2$ (300K)&873&2.41&2.75\\
			ZrSe$_2$ 2\% (300K)&1250&2.28&2.81\\
			ZrSe$_2$ 4\% (300K)&1530&2.05&2.85\\
			ZrSe$_2$ 8\% (300K)&1560&1.27&3.66\\
			TiSe$_2$ (300K)&43.7&2.64&1.45\\
			TiSe$_2$ 2\% (300K)&63.8&2.55&1.55\\
			TiSe$_2$ 4\% (300K)&206&2.40&2.72\\
			TiSe$_2$ 8\% (300K)&584&1.66&2.80\\
			HfSe$_2$ (1200K)&329&2.15&9.81\\
			HfSe$_2$ 2\% (1200K)&422&1.80&9.84\\
			HfSe$_2$ 4\% (1200K)&494&1.67&9.86\\
			HfSe$_2$ 8\% (1200K)&531&1.25&9.83\\
			ZrSe$_2$ (1200K)&274&2.31&9.11\\
			ZrSe$_2$ 2\% (1200K)&360&2.13&8.83\\
			ZrSe$_2$ 4\% (1200K)&424&1.82&7.91\\
			ZrSe$_2$ 8\% (1200K)&449&1.23&7.81\\
			TiSe$_2$ (1200K)&83.7&2.44&8.08\\
			TiSe$_2$ 2\% (1200K)&81.7&2.36&6.99\\
			TiSe$_2$ 4\% (1200K)&79.1&2.20&6.74\\
			TiSe$_2$ 8\% (1200K)&108&1.59&6.60\\
		\end{tabular}
		\label{straintransporttable}
	\end{table}
	
	Seebeck coefficient is directly correlated with the number of sharp peaks in the DOS, a consequence of degeneracies and near-degeneracies in the band structure. The results show that tensile strain modifies the energy difference between the conduction bands at the near-degeneracy at high-symmetry k-point $M$. When this difference reaches a minimum, the near degeneracy of the conduction band valleys is at its highest, and an increased S is expected. TiSe$_2$ and ZrSe$_2$ exhibit a minimum at 8\% strain, whereas HfSe$_2$ develops a degeneracy at k-point $M$ at 2\%. As expected, TiSe$_2$ and ZrSe$_2$ reach maximum Seebeck at 8\% strain, while HfSe$_2$ reaches an enhanced S at 2\% strain that changes little with further strain. The DOS figures further confirm these results in that the number and steepness of peaks in the DOS, especially around the fermi level, increase substantially with strain. All monolayers see a greatly enhanced Seebeck with the application of strain, where the peak values shift towards higher potentials. These results indicate that tensile strain is highly effective in augmenting the Seebeck coefficient of monolayer materials, and are consistent both with current theory and previous theoretical research on other TMD monolayers \cite{sadeghi2019non}.
	
	Electrical conductivity is generally less affected by the application of strain, maintaining largely the same shapes and trends across chemical potential. However, peak values see decreases at high potentials with increasing strain, especially at high temperatures, while the low-potential region of near-zero conductivity widens progressively, reflecting the widening bandgap.
	
	Power factor is shown to be augmented by strain at low temperatures -- at 300K, an overall enhancement in the performance of all monolayers is caused by the increased Seebeck that comes with strain, where the peak PF of HfSe$_2$ increases by 16\%, that of ZrSe$_2$ increases by 33\%, and that of TiSe$_2$ increases by 93\%. These results show that MSe$_2$ can be substantially improved at low temperatures, bolstering the already promising HfSe$_2$ and ZrSe$_2$ and elevating the semimetal TiSe$_2$ to the performance level of high-performing thermoelectric semiconductors. However, PF at high temperatures sees no improvement and a generally decreasing trend as strain increases due to the rapidly widening zone of near-zero electric conductivity that grows with strain.

	\subsection{MSe$_2$ / MSe$_2$ heterobilayer}
	
	\begin{figure} [H]
		\centering
		\begin{subfigure}[b]{0.3\textwidth}
			\includegraphics[width=\textwidth]{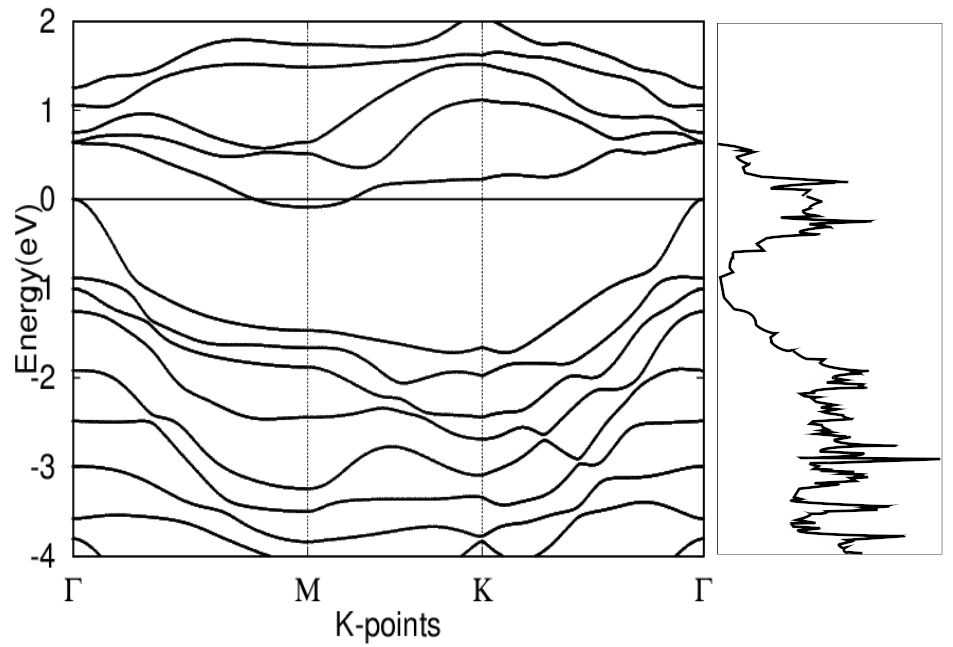}
			\caption{}
			\label{HfTiband}
		\end{subfigure}
		\begin{subfigure}[b]{0.3\textwidth}
			\includegraphics[width=\textwidth]{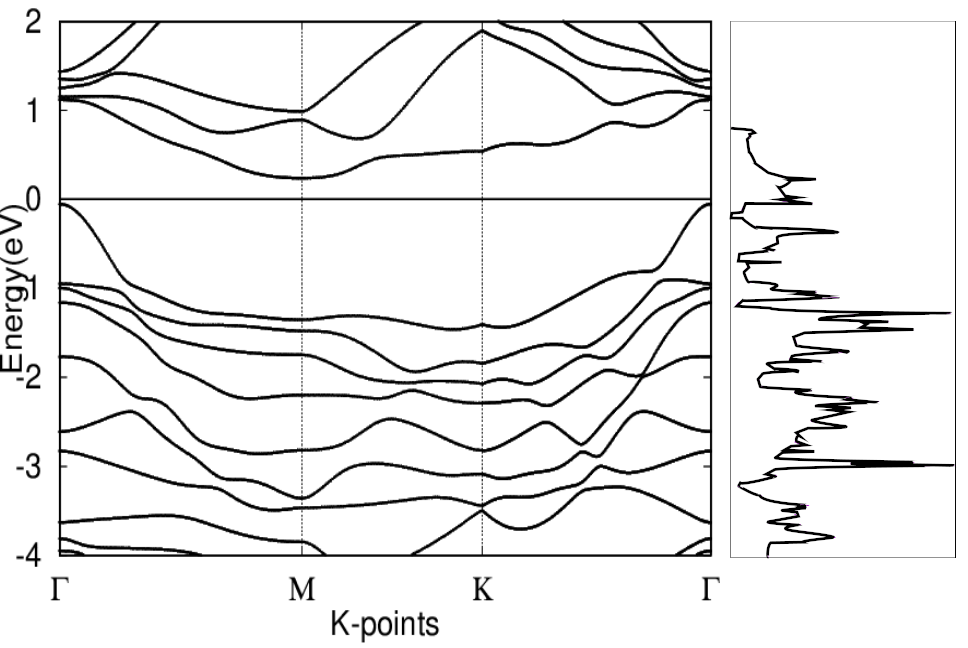}
			\caption{}
			\label{ZrHfband}
		\end{subfigure}
		\begin{subfigure}[b]{0.3\textwidth}
			\includegraphics[width=\textwidth]{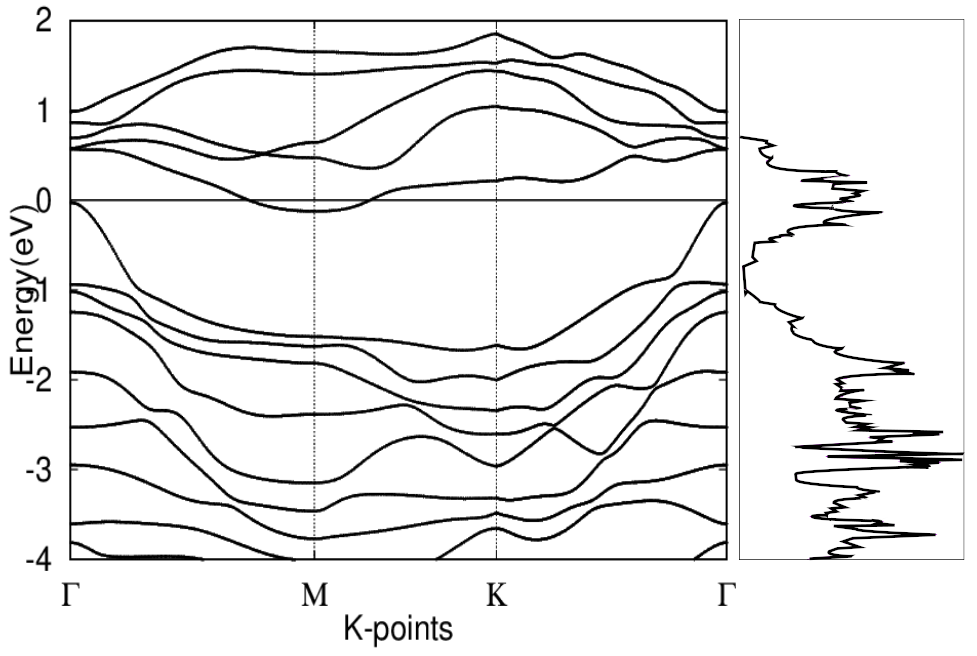}
			\caption{}
			\label{ZrTiband}
		\end{subfigure}
		\label{bilayerband}
		\caption{Band structure and DOS calculated for (a) HfSe$_2$/TiSe$_2$ (b) ZrSe$_2$/HfSe$_2$ (c) ZrSe$_2$/TiSe$_2$}
	\end{figure}
	\begin{table}[hpt]
		\caption{Calculated peak transport values: Seebeck coefficient (S), electrical conductivity ($\sigma/\tau$), power factor (PF) for MSe$_2$/MSe$_2$ heterobilayers at 300K and 1200K}
		\centering
		\begin{tabular}{c c c c}
			Material&S ($\mu$V/K))& $\sigma/\tau$ (10$^{-4}$ $\Omega^{-1}m^{-1}s^{-1}$)&PF (10$^{11}$ W/K$^2$ms)\\
			\hline
			\hline
			HfSe$_2$ / TiSe$_2$ (300K)&41.5&1.78&0.71\\
			ZrSe$_2$ / HfSe$_2$ (300K)&319&1.32&5.60\\
			ZrSe$_2$/ TiSe$_2$ (300K)&370&1.55&7.47\\
			HfSe$_2$ / TiSe$_2$ (1200K)&78.3&1.64&3.21\\
			ZrSe$_2$ / HfSe$_2$ (1200K)&100&1.09&4.59\\
			ZrSe$_2$/ TiSe$_2$ (1200K)&118&1.32&7.69\\
		\end{tabular}
		\label{bitransporttable}
	\end{table}
	
	Of the heterobilayers, ZrSe$_2$/HfSe$_2$ is an indirect band-gap semiconductor, and HfSe$_2$/TiSe$_2$ and ZrSe$_2$/TiSe$_2$ are semi-metals, indicating mixing of the properties of the corresponding monolayers. In any case, all heterobilayers exhibit a much lower band-gap nature than their semiconducting monolayer counterparts.
	
	\begin{figure}[H]
		\centering
		\includegraphics[width=\textwidth]{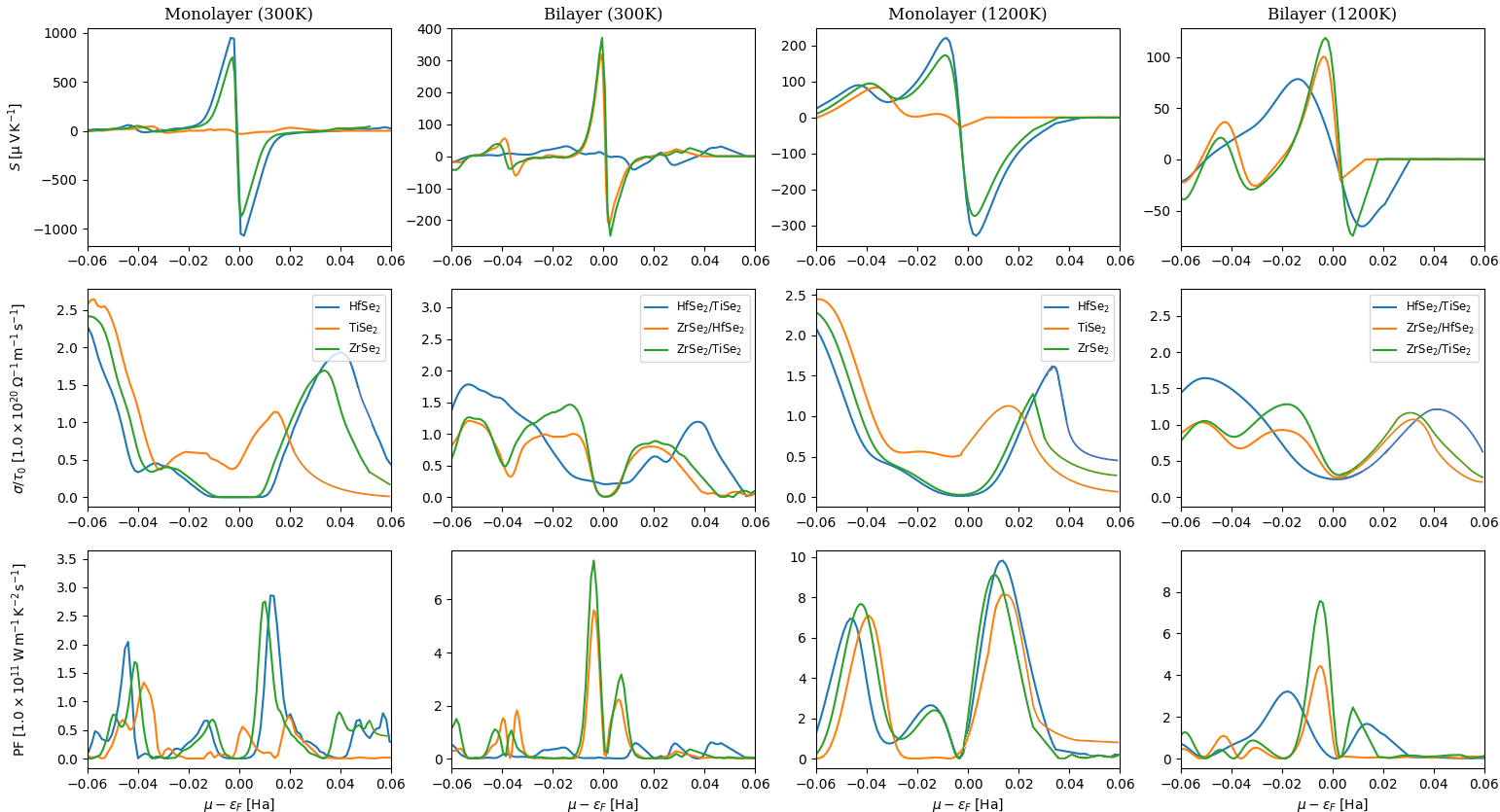}
		\label{Bitransport}
		\caption{Thermoelectric properties: Seebeck coefficient (S), electrical conductivity ($\sigma/\tau$), power factor (PF) plotted as a function of chemical potential at 300K and 1200K for: monolayer and bilayer MSe$_2$ (M = Ti, Zr, Hf) }
	\end{figure}	
	
	Consistent with the decreased band-gap and more metallic nature, heterobilayers have lower Seebeck than monolayers. Similar to the monolayer, heterobilayers show greatest Seebeck in regions of low chemical potential, although a higher value is seen in the p-type region than in the n-type region. Once more, the Seebeck values drop precipitously with increased temperature by about the same amount as was observed in the case of the monolayer, falling to about a fourth of the value.
	
	Heterobilayers exhibit an electrical conductivity that changes much less with chemical potential, and has higher values at low potential. Peak values, however, are lower than their monolayer counterparts. Differing from the behavior of the monolayer, an increase in temperature leaves the electrical conductivity unchanged; whereas the monolayer's peak values shift towards lower potentials at high temperatures, the peak values shown in the heterobilayer remain in the same locations. This variation from the monolayer data is caused by band engineering: the general resistance to change reflects the more metallic state of the heterobilayers.
	
	Each heterobilayer exhibits higher PF than its monolayer constituents, thanks to an increased electrical conductivity at low potential combined with a relatively unchanged Seebeck at the point of peak. As with the monolayer, PF generally increases with temperature; however, the increase is much less marked. This is due to the electrical conductivity, which changes much less with temperature than in the case of the monolayer.
	
	Compared with the corresponding monolayers, the heterobilayers ZrSe$_2$/HfSe$_2$ and ZrSe$_2$/TiSe$_2$ perform significantly better at low temperatures, outperforming all monolayers by two to three times, whereas at high temperatures, the thermoelectric performance of all bilayers is worse. This is largely because of an auspicious combination of a widened Seebeck curve and a narrowed electrical conductivity curve in the monolayer, where no such combination occurs in the bilayer.
	
	\begin{table}[H]
		\caption{Calculated peak Power factor (PF) values of all MSe$_2$ configurations at 300K and 1200K}
		\centering
		\begin{tabular}{c c c}
			Material&PF (10$^{11}$ W/K$^2$ms) 300K&PF (10$^{11}$ W/K$^2$ms) 1200K\\
			\hline
			\hline
			TiSe$_2$&1.45&8.08\\
			HfSe$_2$&2.92&9.81\\
			ZrSe$_2$&2.75&9.11\\
			TiSe$_2$ (strained)&2.80&6.99\\
			HfSe$_2$ (strained)&3.40&9.86\\
			ZrSe$_2$ (strained)&3.66&8.83\\
			HfSe$_2$/TiSe$_2$&0.71&3.21\\
			ZrSe$_2$/HfSe$_2$&5.60&4.59\\
			ZrSe$_2$/TiSe$_2$&7.47&7.69\\
		\end{tabular}
		\label{PFTable}
	\end{table}
	
	Among pure monolayer materials, HfSe$_2$ performed highest due to greater valence band near-degeneracy and higher variance in the density of states near band edges and the fermi level, whereas strained monolayer ZrSe$_2$ achieves similar near degeneracy and therefore similar thermoelectric performance.
	
	Although (semi) metals typically have lower thermoelectric transport performance than semiconductors, the highest performing heterostructure was the semiconductor/semimetal combination ZrSe$_2$/TiSe$_2$, due to an auspicious combination of increased intrinsic electrical conductivity and relatively preserved Seebeck. This is to be compared to the semiconductor / semiconductor heterostructure ZrSe$_2$ / HfSe$_2$, where intrinsic electrical conductivity is very nearly unchanged due to both of its constituent materials having near-zero intrinsic conductivities. This result indicates that interestingly, combining a promising pure semiconducting thermoelectric monolayer with a metal one can boost thermoelectric performance more than combination with semiconductor that, when isolated, performs even better than the first.
	
	All of the monolayers are seen to be promising thermoelectric materials, especially at high temperatures, while The ZrSe$_2$ heterobilayers, especially ZrSe$_2$ / TiSe$_2$, are shown to be materials of remarkable thermoelectric power at low temperatures. These observations provide for the application and creation of thermoelectric materials which perform at high levels in both low and high temperature conditions.

	\section{Conclusion}
	
	In summary, we use density functional theory and semi-classical Boltzmann transport calculations to explore various configurations of the low dimension TMD materials MSe$_2$ (M = Zr,Hf,Ti) in search of especially promising thermoelectric materials with high PF at both low and high temperatures. We tested the effects of biaxial tensile strain by applying varying discrete strains ($\epsilon = 2\%, 4\%, 8\%$). Our results indicated that strain was highly effective in enhancing Seebeck and thermopower, especially at low temperatures -- at 300K, PF was enhanced up to 93\% by the application of strain. Strained ZrSe$_2$ was shown to perform best at low temperatures, while strained HfSe$_2$ was shown to perform best at higher temperatures. Finally, we explored the properties of the relevant van der Waals heterobilayers. Transport properties (Seebeck, electrical conductivity, power factor) were calculated as a function of chemical potential at 300K and 1200K, revealing remarkably enhanced PF at low temperatures, with ZrSe$_2$/TiSe$_2$ performing best and exhibiting a 415\%/171\% increased PF over its component monolayers at 300K. These results make for highly promising applications in renewable thermoelectric applications. 
	
	\section{Acknowledgments}
	
	We would like to thank Dr. Gefei Qian for technical support.
	
	\newpage
	\bibliographystyle{elsarticle-harv}
	\bibliography{references.bib}{}

\end{document}